\documentclass[preprint]{elsarticle}

\usepackage{natbib} 
\usepackage{verbatim}
\usepackage{graphicx} 
\usepackage{array} 
\usepackage{paralist} 
\usepackage{verbatim} 
\usepackage{pdflscape} 
\usepackage{amsmath} 
\usepackage{float} 
\usepackage{algorithm}
\usepackage{algorithmic}
\usepackage{amssymb}
\usepackage{epstopdf}
\usepackage{pdflscape}

\usepackage{algorithmic}
\usepackage{verbatim}
\usepackage{enumerate}
\usepackage{verbatim}
\usepackage{enumerate}
\usepackage{arydshln}
\usepackage[T1]{fontenc}
\usepackage[utf8]{inputenc}
\usepackage{geometry}
\usepackage{caption}
\usepackage{subcaption}
\usepackage{tikz}
\usetikzlibrary{shapes,arrows}

\begin{document}
\begin{frontmatter}

\title{Circuit Synthesis of Electrochemical Supercapacitor Models}

\author[a]{Ross Drummond }
\author[a]{Shi Zhao }
\author[a]{David A. Howey }
\author[a]{Stephen R. Duncan \corref{mycorrespondingauthor}}

\cortext[mycorrespondingauthor]{Corresponding author. Tel +44 1865 283261. Fax 1865 273906}
\ead{stephen.duncan@eng.ox.ac.uk}

\address[a]{Department of Engineering Science, University of Oxford, Oxford, UK, OX1 3PJ}

\begin{abstract}
This paper is concerned with the synthesis of RC electrical circuits from physics-based supercapacitor models describing conservation and diffusion relationships. The proposed synthesis procedure uses model discretisation, linearisation, balanced model order reduction and passive network synthesis to form the circuits. Circuits with different topologies are synthesized from several physical models. This work will give greater understanding to the physical interpretation of electrical circuits and will enable the development of more generalised circuits, since the synthesized impedance functions are generated by considering the physics, not from experimental fitting which may ignore certain dynamics.
\end{abstract}


\end{frontmatter}

\section{Introduction}

Electric double layer (EDL) supercapacitors are electrical energy storage devices for high power applications that store energy electrostatically on porous electrodes with high specific surface areas \cite{lu2013supercapacitors}. Since, the energy storage mechanism of supercapacitors does not involve chemical reactions, supercapacitors exhibit higher power densities, lower energy densities and are less susceptible to temperature and degradation effects than the high energy devices \cite{luo2015overview}. This means that supercapacitors can be considered to be hybrid energy storage devices, sitting between the high power devices like dielectric capacitors with the high energy devices on the Ragone plot \cite{winter2004batteries}. This unique set of energy storage characteristics has led to supercapacitors being successfully applied in a range of applications including hybrid electric vehicles, fault ride through and load levelling \cite{schneuwly2000properties}.

In the fields of science and engineering, progress is made through the development of dynamic models that describe physical systems. Amongst other benefits, models allow predictions to be made about the physical system and can be incorporated within a control system for improved performance. Mathematical models of supercapacitors can be generalised into three subgroups: equivalent circuits (EC), artificial neural networks (ANN) and physics based (PB) models. In this paper, EC circuits and PB models of EDL supercapacitors typically with current as input and voltage as output will be considered. In an equivalent circuit model, the electrical behaviour of the supercapacitors is approximated by fitting a passive RC circuit \cite{devillers2014review}. The resulting model equations are, in general, linear ordinary differential equations (ODEs), which are relatively straightforward to solve. This simplicity has led to equivalent circuit models being the dominant modelling technique in this field, especially when the model is used for simulation or real time estimation, not for design purposes. However, this simplicity comes at a cost, as the model states have unclear physical meaning and so no knowledge is gained about the internal state of the system. Also, the fitting procedure of EC models means that they are only suitable around some operating condition and have to be re-parameterised when the system deviates significantly from this point. Numerous EC models have been presented in the literature \cite{rafik2007frequency, spyker2000classical,buller2005impedance}, with three common circuits being compared for an electric vehicle application in \cite{zhang2015comparative}.

Physics based (PB) models use a set of partial differential equations (PDEs) to describe charge and mass transport phenomena occurring within the supercapacitor. These PDEs describe physical phenomenon, such as electrolyte diffusion and charge conservation \cite{verbrugge2005microstructural}. Since the underlying equations are PDEs, not ODEs, PB models are more complex to simulate than EC models and are usually implemented with some form of spatial discretisation technique, such as the finite difference method \cite{verbrugge2005microstructural}. In \cite{verbrugge2005microstructural}, the physical PDE equations were established and shown to match experimental data fairly well. Several studies have subsequently expanded upon this PB model, for example, by studying electrode 3D effects and parameter sensitivities \cite{allu2014generalized}, the computational implementation with an efficient spectral collocation discretisation \cite{drummond2015low}, implementation with the multi-physics modelling programme COMSOL \cite{madabattulamodeling}, a reduced order PDE system where concentration effects were ignored \cite{romero2010reduced}, the inclusion of temperature effects \cite{d2014first} and analytic solutions for constant current and electrical impedance spectroscopy charging profiles are given in \cite{srinivasan1999mathematical}. Physics based models have also been developed for related electrical energy storage devices. For example, the Newman model \cite{doyle1993modeling} has been widely studied for lithium-ion batteries, both in terms of its implementation \cite{bizeray2013advanced} and it's incorporation within a control system \cite{chaturvedi2010algorithms}.

Even though both EC and PB supercapacitor models describe the same physical device, there has not been any real overlap between the two methods and they are often treated as two separate approaches. Those efforts which have been made to link the approaches tend to give a qualitative, rather than quantitative, relationship between the two. The purpose of this paper is to bridge this gap by linking the two approaches using a mathematical transformation, such that equivalent circuits can be synthesized from the physical PDEs. This mathematical transformation will use the tools of balanced model order reduction \cite{dullerud2000course} and passive network synthesis \cite{guillemin1957synthesis}. In order for circuits to be realised, it is necessary to give a state-space realisation of the impedance function.  State-space realisations of analytic impedance functions by means of model order reduction and Taylor series expansion for PB lithium ion models was studied by Smith et al in \cite{smith2007control,jun2015state} and the work of this paper is concerned with the circuit synthesis of similar realisations. The work of this paper could be said to generalise \cite{rael2013using}, where a specific circuit is designed to analogously describe a PB lithium ion model, since the goal of this paper is show that a wide \textit{class} of  PB models can be synthesised by a wide \textit{class} of circuits. The circuits developed by this approach will have a physical basis and should be more robust than those which fit an impedance function to data, as is commonly adopted.  It should be noted that this work is not strictly focused on model development, but instead on PB model analysis in terms of electrical components.


The paper is structured as follows; in Section 1, several PB and EC models are introduced. The physics based model is linearised in Section 2 and in Section 3, the tools of passive network synthesis and model order reduction are used to form circuits from the PB model. 


\section{Models}
\begin{figure}
\centering
\graphicspath{ {./Figures/} }
\includegraphics[width=0.6\columnwidth]{{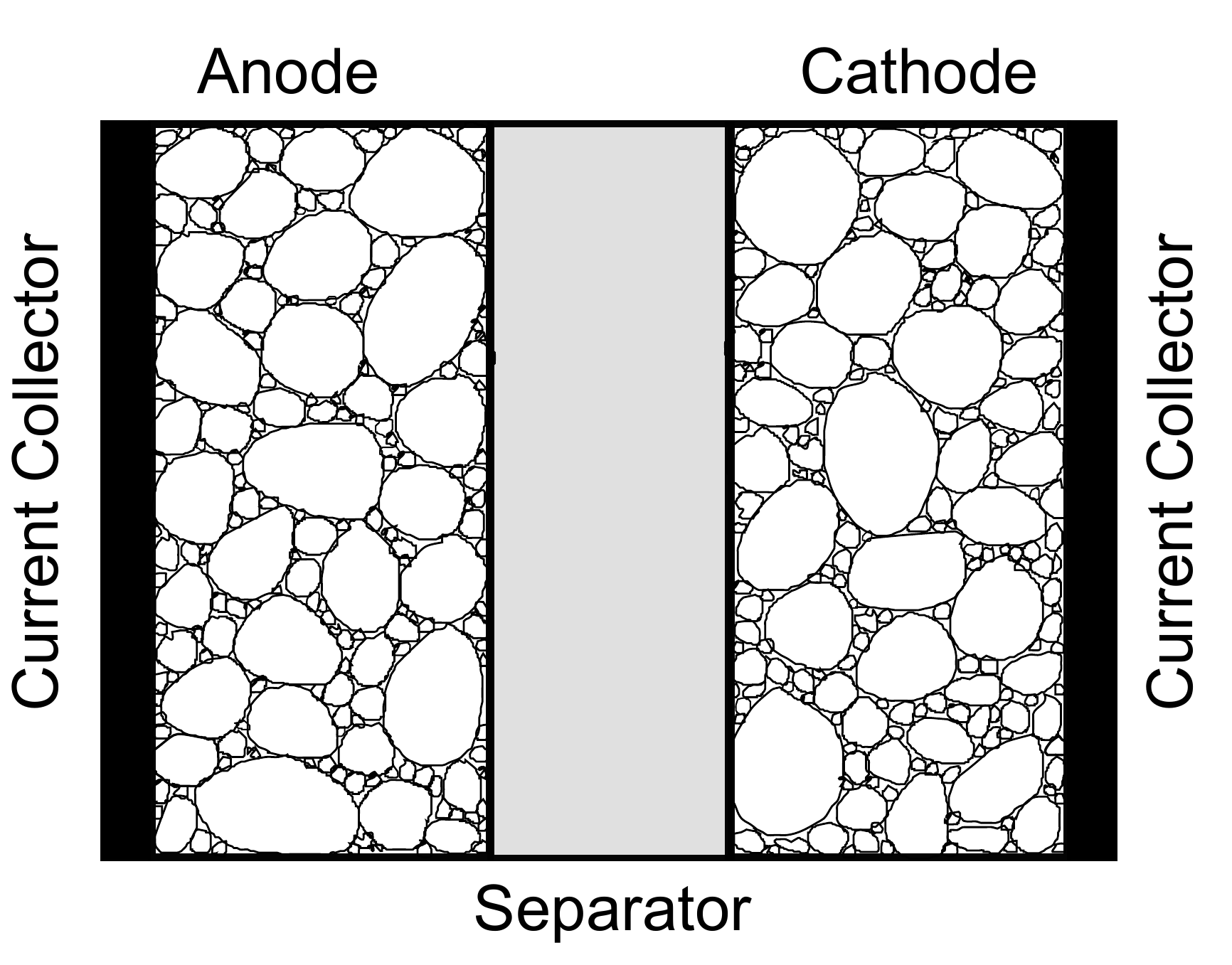}}
\caption{The standard construction of a supercapacitor.}
\label{fig:Supercap}
\end{figure}

The PB supercapacitor model which will be studied in this paper was developed in \cite{drummond2015low} and is a reformulation of the model set out in  \cite{verbrugge2005microstructural}. For the purposes of this paper, this PB model is treated as a `true' model which describes the whole dynamics of the device and was shown to match up with experimental data \cite{verbrugge2005microstructural}. The model has three domains, one for each electrode and one for the separator, with the electrically insulating separator preventing a short circuit. In order for the model to be tractable, several assumptions have to be introduced, as outlined in \cite{verbrugge2005microstructural}. These include the homogenisation of the electrode  structure using porous electrode theory and fixing the capacitance as a lumped parameter, even though capacitance has been shown to change with variables such as the voltage \cite{Wu20147885}. The boundary conditions of the model are also outlined in \cite{verbrugge2005microstructural} and are applied at the separator/electrode and current collector/electrode interfaces. These boundary conditions can be summarised as enforcing conservation of ionic flux and current. The current collectors are responsible for the transfer of current to and from the system. This setup is shown in Fig. \ref{fig:Supercap}.

The three partial differential algebraic equations of the PB model describe:
	
\begin{itemize}
  \item Charge conservation across the double layer
\end{itemize}	
\begin{equation}\label{eq:2}
aC \frac{\partial (\phi_1-\phi_2)}{\partial t}= \sigma \frac{\partial^2 \phi_1}{\partial x^2}
\end{equation} 
\begin{itemize}
  \item Elctrolyte diffusion
\end{itemize}	
\begin{equation}\label{eq:1}
\epsilon \frac{\partial c}{\partial t}= D \frac{\partial^2 c}{\partial x^2}-\frac{aC}{F}\Bigg( t_- \frac{dq_+}{dq}+t_+\frac{dq_-}{dq} \Bigg) \frac{\partial (\phi_1-\phi_2)}{\partial t},
\end{equation}
\begin{itemize}
  \item Ohm's Law
\end{itemize}	
\begin{equation}\label{eq:3}
 \kappa \bigg(\frac{RT(t_+-t_-)}{F}\bigg) \frac{\partial}{\partial x}\text{ln }(c)
+ \sigma \frac{\partial (\phi_1- \phi_2 ) }{\partial x}  
+  \bigg(\kappa \frac{\partial}{\partial x}+\sigma \frac{\partial}{\partial x}\bigg) \phi_2
+i = 0
\end{equation}
with specific capacitance $aC$, potential in the electrode $\phi_1$, potential in the electrolyte $\phi_2$, electrode conductivity $\sigma$, porosity $\epsilon$, diffusion constant $D$, Faraday constant $F$, transference numbers $t_+$ and $t_-$, $\frac{dq_{+/-}}{dq}$ describing the change in surface concentration of an ion associated with a change in the surface charge on the electrode $q$, electrolyte conductivity $\kappa$, gas constant $R$, temperature $T$ and current density $i$. The values of these parameters used in this model are given in Table \ref{tab:GlobalParams}. The output of the model $y$ is the voltage $V$
\begin{equation}
y = V = \phi_1 |_{x = L} - \phi_1|_{x = 0}.
\end{equation}
In the electrodes, equations (\ref{eq:2}),  (\ref{eq:1}) and (\ref{eq:3}) have state-space form
\begin{align}\label{state_space}
\begin{split}
&\begin{bmatrix}\epsilon & \frac{aC}{F}(t_-\frac{dq_+}{dq}+t_+\frac{dq_-}{dq}) & 0\\
0 & aC & 0 \\
0 & 0 & 0 \end{bmatrix}
\begin{bmatrix} \dot{c} \\ \dot{\phi}_1-\dot{\phi}_2 \\ \dot{\phi}_2\end{bmatrix}
= 
\begin{bmatrix} 0 \\0 \\ \kappa \bigg(\frac{RT(t_+-t_-)}{F}\bigg) \frac{\partial}{\partial x}\end{bmatrix}\text{ln }(c) 
\\
& \qquad + \begin{bmatrix}D \frac{\partial^2}{\partial x^2} & 0 & 0\\
0 &\sigma \frac{\partial^2}{\partial x^2}&\sigma \frac{\partial^2}{\partial x^2} \\
0& \sigma \frac{\partial}{\partial x} & \kappa \frac{\partial}{\partial x}+ \sigma \frac{\partial}{\partial x} \end{bmatrix}
\begin{bmatrix} c \\ \phi_1- \phi_2 \\ \phi_2\end{bmatrix} 
+\begin{bmatrix}0 \\0 \\ 1 \end{bmatrix}i
\end{split}
\end{align}
and in the separator 
\begin{gather}\label{eqn:seperator_state}
\begin{aligned}
\begin{bmatrix}\epsilon & 0 \\
0 & 0 \\
\end{bmatrix}
\begin{bmatrix} \dot{c} \\ \dot{\phi}_2 \\\end{bmatrix}
= 
& \begin{bmatrix} D \frac{\partial^2}{\partial x^2} & 0 \\
0 & \kappa \frac{\partial}{\partial x}\end{bmatrix}
\begin{bmatrix} c \\ \phi_2 \end{bmatrix}
+ \begin{bmatrix} 0 \\ \kappa \big(\frac{t_+-t_-}{f}\big) \frac{\partial}{\partial x}\end{bmatrix}\text{ln } (c) +
\begin{bmatrix}0 \\ 1 \end{bmatrix}i.
\end{aligned}
\end{gather} 

In implementation, discretisation methods are normally applied to the spatial differentiation operators of partial differential algebraic equations (DAE) systems such as (\ref{state_space}). Discretisation gives a finite dimensional approximation to the infinite dimensional PDE problem, resulting in a significantly simpler problem to solve. Upon discretisation, the spatial derivative operator $\partial / \partial x$ is approximated by a differentiation matrix $\hat{\mathbf{D}}_s$, with the subscript $s$ implying that the matrix accounts for state $s$'s boundary conditions. These boundary conditions are enforced by the patching technique of \cite{trefethen2000spectral}. As outlined in \cite{drummond2015low}, the spectral collocation method will be used to discretise the model equations as this results in lower order models for a given level of solution accuracy \cite{trefethen2000spectral}. The discretised version of the electrode state-space system (\ref{state_space}) is
\begin{align}\label{discrete_state_space_3}
\begin{split}
&\begin{bmatrix}\epsilon & \frac{aC}{F}(t_-\frac{dq_+}{dq}+t_+\frac{dq_-}{dq}) & 0\\
0 & aC & 0 \\
0 & 0 & 0 \end{bmatrix}
\begin{bmatrix} \mathbf{\dot{c}} \\ \mathbf{\dot{\phi}_1}-\mathbf{\dot{\phi}_2}\\ \mathbf{\dot{\phi}_2}\end{bmatrix}
 = \begin{bmatrix} 0 \\0 \\ \kappa \bigg(\frac{RT(t_+-t_-)}{F}\bigg) \hat{\mathbf{D}}_{\text{ln }c}\end{bmatrix}\text{ln } ( \mathbf{c})
\\
&
\qquad \begin{bmatrix}D \hat{\mathbf{D}}^2_c & 0 & 0\\
0 &\sigma \hat{\mathbf{D}}^2_{\phi_1}&\sigma \hat{\mathbf{D}}^2_{\phi_1} \\
0& \sigma \hat{\mathbf{D}}_{\phi_1} & \kappa \hat{\mathbf{D}}_{\phi_2}+\sigma \hat{\mathbf{D}}_{\phi_1} \end{bmatrix}
\begin{bmatrix} \mathbf{c} \\ \mathbf{\phi_1}- \mathbf{\phi_2 }\\ \mathbf{\phi_2}\end{bmatrix} 
 +
\begin{bmatrix}0 \\ 0 \\ 1 \end{bmatrix}i,
\end{split}
\end{align}
and, similarly, for the separator
\begin{gather}\label{eqn:seperator_state_discrete}
\begin{aligned}
\begin{bmatrix}\epsilon & 0 \\
0 & 0 \\
\end{bmatrix}
\begin{bmatrix} \dot{\mathbf{c}} \\ \dot{\mathbf{\phi_2}} \\\end{bmatrix}
= 
& \begin{bmatrix} D \hat{\mathbf{D}}^2_c & 0 \\
0 & \kappa \hat{\mathbf{D}}^2_{\phi_2}\end{bmatrix}
\begin{bmatrix} \mathbf{c} \\ \mathbf{\phi_2} \end{bmatrix}
+& \begin{bmatrix} 0 \\ \kappa \big(\frac{RT(t_+-t_-)}{F}\big) \hat{\mathbf{D}}_{\text{ln }c}\end{bmatrix}\text{ln }  (\mathbf{c}) +
\begin{bmatrix}0 \\ 1 \end{bmatrix}i.
\end{aligned}
\end{gather}
In the following analysis, only (\ref{discrete_state_space_3}) will be studied as (\ref{eqn:seperator_state_discrete}) can be embedded into the structure of (\ref{discrete_state_space_3}) by expanding the dimension of the model states. The discretised version of the model output is
\begin{equation}\label{output1}
y = \begin{bmatrix} 0 & C_1\end{bmatrix}\begin{bmatrix}\mathbf{c} \\ \mathbf{\phi_1}-\mathbf{\phi_2}\end{bmatrix} + C_1 \mathbf{\phi_2}+ D_1 i.
\end{equation}

It is possible to convert the DAE (\ref{discrete_state_space_3}) into an ODE by solving the algebraic equation
\begin{equation}\label{algebraic2}
 \kappa \bigg(\frac{t_+-t_-}{f}\bigg) \hat{\mathbf{D}}_{\text{ln }c}\text{ln } (\mathbf{c} )
+ \sigma \hat{\mathbf{D}}_{\phi_1}  (\mathbf{\phi_1}- \mathbf{\phi_2 }) 
+  \bigg(\kappa \hat{\mathbf{D}}_{\phi_2}+\sigma \hat{\mathbf{D}}_{\phi_1}\bigg) \mathbf{\phi_2} 
+i = 0
\end{equation}
for the algebraic variable $\phi_2$
\begin{equation}\label{algebraic3}
 \mathbf{\phi_2}  = -   \left(\kappa \hat{\mathbf{D}}_{\phi_2}+\sigma \hat{\mathbf{D}}_{\phi_1}\right) ^{-1}
\left( \kappa \bigg(\frac{t_+-t_-}{f}\bigg) \hat{\mathbf{D}}_{\text{ln }c}\text{ln } (\mathbf{c} )
+ \sigma \hat{\mathbf{D}}_{\phi_1}  (\mathbf{\phi_1}- \mathbf{\phi_2 }) 
+i \right)
\end{equation}
 This solution exists if one of the potentials is set as a reference. Since only the potential difference, not the actual potential values themselves, is important, this reference can be enforced. If no reference is set, then the range of potentials which would give the same potential difference would be infinite and there would be an infinite number of solutions to (\ref{algebraic3}). Reformulating (\ref{discrete_state_space_3}) using (\ref{algebraic3}) gives the following equation system 
\begin{subequations}\label{trunc2}
\begin{align} \label{trunc}
\hspace{-1cm} \begin{bmatrix}M_{11} & M_{12} \\ 0 & M_{22}\end{bmatrix}
\begin{bmatrix} \mathbf{\dot{c}} \\ \mathbf{\dot{\phi}_1}-\mathbf{\dot{\phi}_2}\end{bmatrix}
&= 
\begin{bmatrix} \hat{A}_{11} & 0 \\ 0 & \hat{A}_{22}\end{bmatrix}
\begin{bmatrix}\mathbf{c} \\ \mathbf{\phi_1}-\mathbf{\phi_2}\end{bmatrix} +
\begin{bmatrix} 0 \\ B_1\end{bmatrix} \text{ ln }(\mathbf{c} )
+
\begin{bmatrix} 0 \\ B_2\end{bmatrix}i
\end{align}
\begin{align}\label{output}
y = \begin{bmatrix} 0 & \tilde{C}\end{bmatrix}
\begin{bmatrix}\mathbf{c} \\ \mathbf{\phi_1}-\mathbf{\phi_2}\end{bmatrix}  
+ \tilde{D}_1 \text{ ln }(\mathbf{c} )
+
\tilde{D}_2 i
\end{align}
\end{subequations}
whose trajectories evolve along the manifold defined by (\ref{algebraic2}). The general form of (\ref{trunc2}) is 
\begin{subequations}
\begin{align}
M\dot{x} &= Ax + \tilde{B}_1 \text{ ln } (c ) + \tilde{B}_2 i \label{mass}
\\
y &= Cx + \tilde{D}_1 \text{ ln } (c)  + \tilde{D}_2 i \label{output_ode}
\end{align}
\end{subequations}
with state  $x:= [\mathbf{c}^T  ,~ \mathbf{\phi_1}^T- \mathbf{\phi_2}^T]^T$ where $c \in \mathbb{R}^n_+$ and $ \phi_1-\phi_2 \in \mathbb{R}^n$. By inverting the ``mass'' matrix $M$, (\ref{trunc}) can be written as a standard dynamic system
\begin{align}\label{physics}
\dot{x} &= A_mx + B_{1,m} \text{ ln } (c)  + B_{2,m} i.
\end{align}

This paper is concerned with the synthesis of equivalent circuits from physical models such as (\ref{physics}). In the literature, a vast array of equivalent circuit models can be found, thus,  to make the circuit synthesis problem tractable, the class of circuits which will be considered is restricted to the \textit{classical}, \textit{ladder} and \textit{dynamic} circuits which have seen widespread application and are compared in \cite{zhang2015comparative}.

\begin{figure}
\centering
\graphicspath{ {./Figures/} }
\includegraphics[width=0.75\columnwidth]{{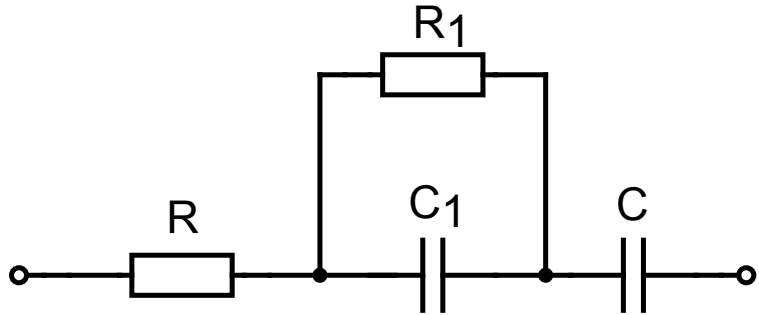}}
\caption{Classical circuit.}
\label{fig:Classical}
\end{figure}

\begin{figure}
\centering
\graphicspath{ {./Figures/} }
\includegraphics[width=0.75\columnwidth]{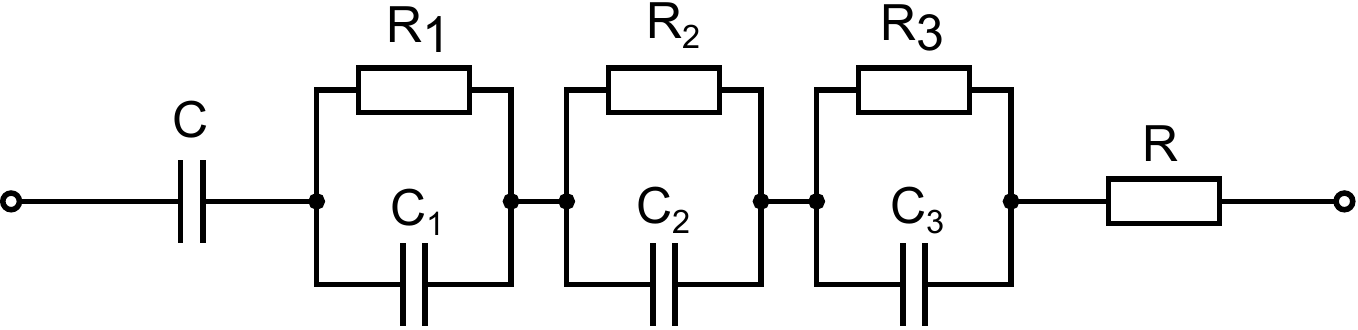}
\caption{Dynamic circuit.}
\label{fig:Dynamic}
\end{figure}

\begin{figure}
\centering
\graphicspath{ {./Figures/} }
\includegraphics[width=0.7\columnwidth]{{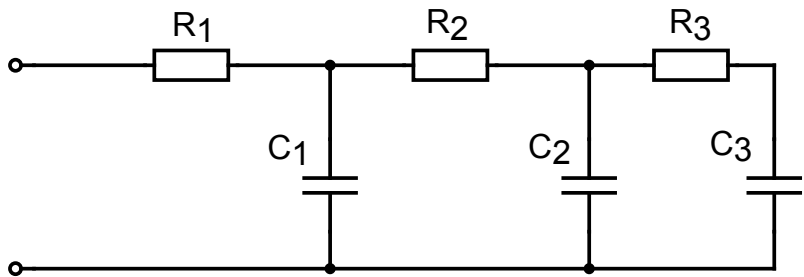}}
\caption{Ladder circuit.}
\label{fig:Ladder}
\end{figure}

 The \textit{classical} model with added series capacitance term, shown in Fig \ref{fig:Classical}, is the simplest model and has dynamics
\begin{subequations}\label{classical}
\begin{align}\label{classical2}
\begin{bmatrix}\dot{x}_1 \\ \dot{x}_2 \end{bmatrix}
&= 
\begin{bmatrix}
0 & 0  \\
0 & -\frac{1}{R_1C_1}  \\
\end{bmatrix}
\begin{bmatrix}x_1 \\ x_2 \end{bmatrix}
+
\begin{bmatrix}\frac{1}{C}\\ \frac{1}{C_1} \end{bmatrix}i
\\
V &= x_1 + x_2 +  Ri.
\end{align}
\end{subequations}
The series capacitance term was added to accommodate the integrators of the PB model. Adding additional time constants to this circuit by the addition of more RC branches results in the \textit{dynamic} circuit, shown in Fig \ref{fig:Dynamic}, with dynamics
\begin{subequations}\label{dynamic}
\begin{align}\label{dynamic2}
\begin{bmatrix}\dot{x}_1 \\ \dot{x}_2 \\ \dot{x}_3 \\ \dot{x}_4\end{bmatrix}
&= 
\begin{bmatrix}
0 & 0 & 0 & 0 \\
0 & -\frac{1}{R_1C_1} & 0& 0 \\
0 &0& -\frac{1}{R_2 C_2} & 0 \\
0 & 0 &0& -\frac{1}{R_3 C_3} 
\end{bmatrix}
\begin{bmatrix}x_1 \\ x_2 \\ x_3 \\ x_4\end{bmatrix}
+
\begin{bmatrix}\frac{1}{C}\\ \frac{1}{C_1} \\ \frac{1}{C_2} \\ \frac{1}{C_3}\end{bmatrix}i
\\
V &= x_1 + x_2 + x_3  + x_4 + R_si.
\end{align}
\end{subequations}
The third circuit which will be studied is the \textit{ladder} circuit of Fig \ref{fig:Ladder}
\begin{subequations}\label{ladder}
\begin{align}\label{ladder2}
\begin{bmatrix}\dot{x}_1 \\ \dot{x}_2 \\ \dot{x}_3\end{bmatrix}
&= 
\begin{bmatrix}
-\frac{1}{R_2 C_1} & \frac{1}{R_2 C_1} & 0 \\
\frac{1}{R_2 C_2} & -\frac{R_2+R_3}{R_3 R_2 C_2} & \frac{1}{R_3 C_2} \\
0 & \frac{1}{R_3 C_3} & -\frac{1}{R_3 C_3} 
\end{bmatrix}
\begin{bmatrix}x_1 \\ x_2 \\ x_3\end{bmatrix}
+
\begin{bmatrix}\frac{1}{C_1}\\ 0 \\ 0\end{bmatrix}i
\\
V &= x_1 + R_1i.
\end{align}
\end{subequations}
%


\section{Circuit Synthesis}
In order to synthesize the three circuits (\ref{classical}), (\ref{dynamic}) and (\ref{ladder}), the nonlinear PB model (\ref{physics}) has to be linearised around the equilibrium concentration $c_e$, where the concentration has diffused to a flat distribution, resulting in the following dynamics
\begin{subequations}\label{linearised2}
\begin{align} \label{linearised}
\dot{x} &= \big(A_m+ [B_{1,m}/c_e ~ 0] \big)x  + B_{2,m} i
\\
y &= \big(C+ [\tilde{D}_{1}/c_e ~ 0] \big)x  + \tilde{D}_2 i. 
\end{align}
\end{subequations}
The logarithmic nonlinearity of (\ref{physics}) is fairly benign when $c_e \gg 0$ and only becomes significant when the concentration approaches zero. 
The near linear voltage-current dynamics of supercapacitors makes them more suitable for circuit realisation than other electrochemical devices such as lithium ion batteries, which exhibit highly nonlinear behaviour, for example with the Butler-Volner equation \cite{bizeray2013advanced}.  


The dimensionality $n$  of the discretised physical state-space system (\ref{linearised2}) is, in general, greater than the number of RC branches of typical equivalent circuit models, which most commonly have 2 or 3 branches \cite{zhang2015comparative}. In order for the boundary conditions to be implemented in (\ref{linearised2}), it was found that the electrode domain required a minimum of three discretisation elements and the separator required a minimum of two elements.  This meant that the minimum dimensionality for the PB model (\ref{linearised2}) was 9 and, for circuit synthesis, the order of (\ref{linearised2}) needed to be reduced. There are a number of methods for model order reduction \cite{antoulas2001survey} and in this work, the balanced truncation method was implemented \cite{dullerud2000course}. The first stage of this method is to introduce the observability operator $\Psi_o: \mathbb{R}^n \to \mathcal{L}^{e}_2[0,t_1]$ where
\begin{align}
\|y\|_ 2^2 = \langle x_0^* \Psi^*_o,\Psi_o x_0 \rangle
\end{align}
such that the observability grammian $Y_o$ can be defined by
\begin{align}
Y_o = \Psi^*_o\Psi_o = \int^{t_1}_0 e^{A^*\tau}C^*Ce^{A\tau} d \tau.
\end{align}
This self-adjoint operator maps the initial conditions to the outputs lying in the $\mathcal{L}_2^{e}$ Hilbert space. Similarly, the controllability operator $\Psi_c: \mathcal{L}^{e}_2[-t_2,0] \to \mathbb{R}^n$ 
\begin{align}
x_0 =\Psi_c u
\end{align}
defines the controllability grammian $X_c$
\begin{align}
X_c = \Psi^*_c\Psi_c = \int_{-t_2}^0 e^{A^*\tau}B^*Be^{A\tau} d \tau
\end{align}
which maps all inputs $u \in \mathcal{L}_2^{e}[-t_2,0]$ to the initial condition $x_0 \in \mathbb{R}^n$. If a linear system is Hurwitz stable, controllable and observable, with system matrices $ (A,B,C,D)$, then it has a balanced realisation meaning that there exists a transformation matrix $T$ such that the equivalent system $(\tilde{A}, \tilde{B}, \tilde{C}, \tilde{D}) = (T\tilde{A}T^{-1}, T\tilde{B}, \tilde{C}T^{-1},D)$ satisfies
\begin{align}
\tilde{X}_c = \tilde{Y}_o = \Sigma
\end{align}
with diagonal $\Sigma >0$. The states which are least controllable in this equivalent system are also least observable.  Introducing a third operator, the Hankel operator $\Gamma_G = \Psi_o\Psi_c: \mathcal{L}_2^{e}[-t_2,0] \to \mathcal{L}_2^{e}[0,t_1]$, maps the model input to the output in a $\mathcal{L}_2^{e}$ sense. This operator gives a bound for the error between a balanced system $G$ and a reduced order system  $G_r$ 
\begin{align}
\| G - G_r \|_{\infty} \geq \sigma_{r+1}
\end{align}
 where $\sigma_1 \geq \sigma_2 \geq \dots \geq \sigma_r \geq \sigma_{r+1} \geq \dots \geq \sigma_n$ are the singular values of $\Gamma_G$ \cite{dullerud2000course}. In this truncation method, the (balanced) state-space matrices are partitioned by 
\begin{align}
A = \begin{bmatrix} A_{11} & A_{12} \\ A_{21} & A_{22}\end{bmatrix}
\quad
B = \begin{bmatrix} B_{1}  \\ B_{2} \end{bmatrix}
\quad
C = \begin{bmatrix} C_{1} & C_{2} \end{bmatrix}
\end{align}
and the resulting reduced order model $(A_{11}, B_1, C_1,D)$ has frequency domain form
\begin{align}
G_r(s) = C_1(Is-A_{11})^{-1}B_1 +D
\end{align}
which is balanced with Hankel Singular norms  $\sigma_1 \dots \sigma_r$ \cite{dullerud2000course}.  Reducing the order of the system in this manner means that the error of the input/output response of the reduced order model is a function of the model order and can be lower bounded by $\sigma_{r+1}$. 

The requirement that the system matrix $A$ be Hurwitz for the balanced truncation method is violated by (\ref{linearised}) since three of the dynamic modes of this system are integrators. This problem was overcome by removing the integrators from the system, performing the model order reduction on the remaining Hurwitz subsystem and then combining this reduced order system back with the integrators. This means that the size of the reduced order system can never be less than 4, however, in synthesis, the three integrator states combine to form a single lumped capacitor. This is shown by considering the dynamics of the integrator states $x_i \in \mathbb{R}^{3}$
\begin{subequations}
\begin{align}
\dot{x}_i &= B_{int}i\\
y_i &= C_{int}x_i
\end{align}
\end{subequations}
where $B_{int} \in \mathbb{R}^{3 \times 1}, C_{int} \in \mathbb{R}^{1 \times 3}$ such that the integrator capacitance $\mathbf{C}$ is given by
\begin{align}
\mathbf{C} = C_{int}B_{int}.
\end{align}
For this reason, the controllability and observability operators have been defined to operate on the $\mathcal{L}_2^{e}$ space, not the full $\mathcal{L}_2$ space. By defining the temporal domain to be a closed set$[-t_2,t_1]$ instead of the open set $(-\infty, \infty)$ as is used with the $\mathcal{L}_2$ space, then the dynamics of (\ref{linearised}) remain bounded for bounded inputs in finite time \cite{jonsson2001lecture}.

Further to the Hurwitz condition, the system is also required to be both observable and controllable for a balanced realisation. These properties were studied in \cite{drummond2015observability} where it was found that the nonlinear model was fully controllable but lost observability when the two transference numbers, $t_+$ and $t_-$ in (\ref{eq:1}), were equal. The presence of unobservable states with equal transference numbers is because the voltage dynamics could then be described independently of the concentration. This meant that no knowledge about the concentrations could be gained from measurements of the voltage. However, for unequal transference numbers, which occurs in most supercapacitor electrolytes \cite{galinski2006ionic}, the model was shown to be fully observable.

The transfer function $G(s)$ of the reduced order system (\ref{linearised2}) is known as the impedance function, giving the $s$-domain  gain from current to voltage. When (\ref{linearised2}) was reduced to a 6 state system using the parameters from Tab \ref{tab:GlobalParams} the impedance function was 
\begin{align}\label{impedance}
G(s) = \frac{V(s)}{I(s)} = \frac{(s+6.56)(s+1.59)(s+0.29)}  {s(s+5.62)(s+1.4) }.
\end{align}
which is the ratio of the Laplace transforms of voltage to current.

Equivalent circuit realisations of the impedance function (\ref{impedance}) will now be obtained using passive network synthesis. Passive network synthesis was studied extensively in the electrical engineering community in the 1930s \cite{bott1949impedance,brune1931synthesis} with the goal being to understand and simulate complex dynamical systems  using electrical components \cite{cha2000fundamentals}. Due to a lack of advancement, interest in this field died down in the 1960s even though many problems still remained open. Network synthesis has recently witnessed somewhat of a revival due to the application of modern mathematical tools  to the classical methods \cite{chen2008electrical} and its application to mechanical systems through the discovery of the inerter \cite{smith2002synthesis}. 

The key requirement for circuit synthesis of an impedance function is that $G(s)$ must be a positive real function. Positive real functions satisfy:
\begin{enumerate}
  \item $G(s)$ is real for real $s$.
  \item Real $G(s) \geq 0$ for real $s \geq 0$.
\end{enumerate}
Positive realness is equivalent to showing that the system is passive, i.e. non energy generating. Passivity of the supercapacitor model has been shown for both the nonlinear discretised system 
and for the fundamental nonlinear partial differential algebraic equations \cite{drummond2015pdes}. The positive-real conditions were satisfied by (\ref{impedance}). Realising a transfer function in terms of a circuit is equivalent to realising it in terms of a state-space system that has a particular structure. Transfer functions are known to have non-unique state-space representation and this explains the vast array of circuit models that can be found in the literature. By restricting the class of circuits to the \textit{classical}, \textit{dynamic} and \textit{ladder} circuits of \cite{zhang2015comparative}, circuit realisations can be obtained, although, it is pointed out that there exists a much wider class of circuits which could be realised by the proposed approach, but are not considered in this paper. 

The methodology of network synthesis is to expand the positive real impedance function $G(s)$ around some point, such as a pole at $s = 0$. The various components of this expansion can then be realised by passive electrical components such as resistors, capacitors and inductors \cite{wadhwa2007network}. The realisation of the impedance function (\ref{impedance}) in terms of the dynamic circuit (\ref{dynamic}) is known as the Foster form of the first kind \cite{wadhwa2007network} and is obtained by continuously removing a pole of $G(s)$ at $s = 0$. This is akin to performing a partial fraction expansion
\begin{align}
G(s) = \frac{1}{k_0} + \frac{k_1}{s+\sigma_1} + \dots + \frac{k_i}{s+\sigma_i} + \dots + k_{\infty}.
\end{align}
The resistances and capacitances of the circuit can then be obtained by the following rules
\begin{align}
C = \frac{1}{k_0} \quad C_i = \frac{1}{k_i} \quad R_i = \frac{k_i}{\sigma_i} \quad R = k_{\infty}.
\end{align}
By reducing the order of the reduced system to 1, the \textit{classical} circuit of (\ref{classical}) is realised. Another method to obtain the RC components for the \textit{dynamic} circuit would be to take the eigen-decomposition of the reduced order physical system (\ref{linearised2}) and then match up the system matrix coefficients to that of (\ref{dynamic}), since both would have diagonal structures. 

\begin{figure}
\centering
\graphicspath{ {./Figures/} }
\includegraphics[width=0.75\columnwidth]{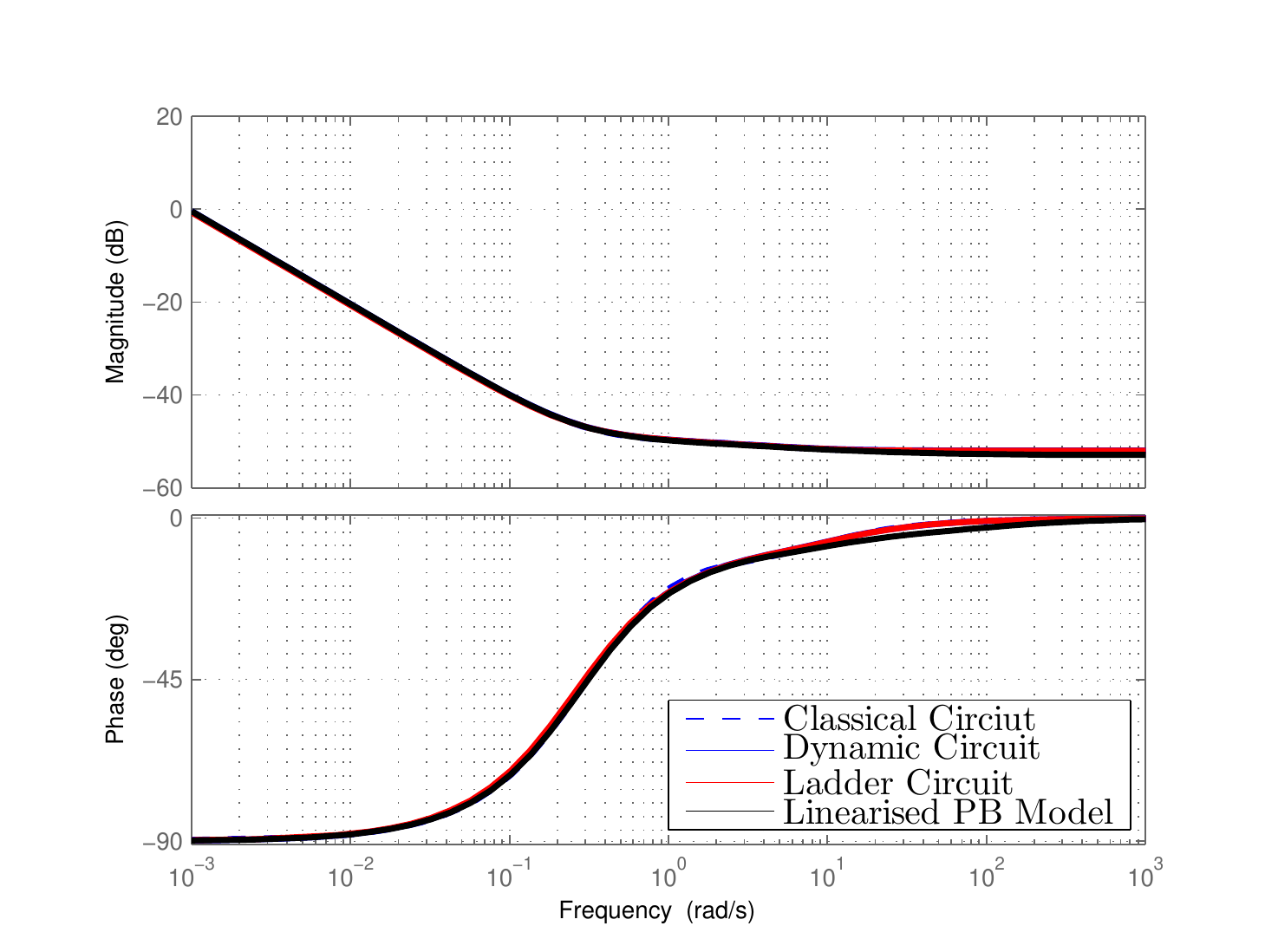}
\caption{Bode plot of the three synthesized circuits and full order linearised physical model.}
\label{fig:bode}
\end{figure}

The \textit{ladder} circuit (\ref{ladder}) can be realised from the impedance function by the continuous removal of poles at $ s =  \infty$. This representation is known as the Cauer form of the first kind and can be obtained by taking a continuous fraction expansion of the impedance function
\begin{equation}
 G(s)  = \alpha_1 + \cfrac{1}{\alpha_2s
          + \cfrac{1}{\alpha_3
	+ \cfrac{1}{\alpha_4 s
          + \cfrac{1}{\begin{matrix}
\alpha_5 +  &  &  \\
 &  \ddots & \\
   & & \alpha_{r-1} + \frac{1}{\alpha_r s}  
 \end{matrix} } } }}.
\end{equation}
 The various resistors and capacitors of the circuit can then be obtained with
\begin{align}
R_i = \alpha_{2i-1}  \quad C_i =  \alpha_ {2i} \quad \text{for } i = 1, \dots,  r/2.
\end{align}
The Cauer form of the second kind of (\ref{impedance}) has the resistors and capacitors swapped around in Fig \ref{fig:Ladder}. The values of the resistors and capacitors for the three circuits are given in Tables \ref{tab:Classical Model}, \ref{tab:Dynamic Model} \&  \ref{tab:Ladder Model}. The Bode plots for the three circuits as well as the full order linearised PB model is shown in Fig \ref{fig:bode}. Since the Bode responses of the circuits match up well with the PB model, this validates the synthesis method. This is to be expected since the circuits are simply reduced order realisations of the physical impedance function.

\begin{figure}
\centering
\graphicspath{ {./Figures/} }
\includegraphics[width=0.75\columnwidth]{{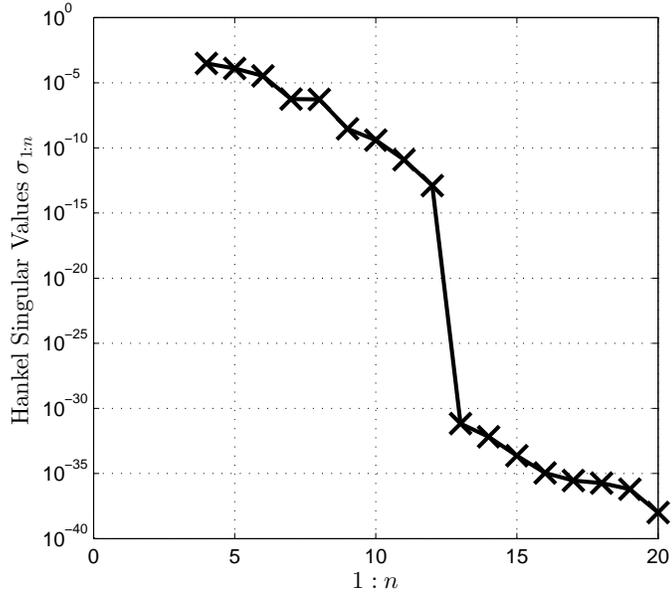}}
\caption{Singular values of the Hankel operator $\Gamma_G$.}
\label{fig:Hankel}
\end{figure}

Using this approach, the errors introduced at each step of the synthesis process (discretisation, linearisation and model order reduction) can be bounded. This gives a quantitative bound for the error of the resulting circuits, when the full order PB model is treated as describing the true dynamics of the system. This contrasts with the rather qualitative errors discussed with fitted EC circuits. The singular values of the Hankel norm describe the error bound for the reduced order system and are shown for a 20 state discretisation of (\ref{linearised2}) in Fig \ref{fig:Hankel}.  The first three singular values in this plot have been removed since they relate to the integrator states. For this discretisation, 12 of the states were related to the concentration $c$ with the remaining 8 being related to the potential difference $\phi_1- \phi_2$. This shows that the concentration states have a greater impact on the input/output dynamics of (\ref{linearised2}) than the potential states. This is because the concentration states evolve by a diffusion process which is much slower than the rapidly decaying fast dynamics of the potentials. This implies that there is no real advantage gained in increasing the number of RC branches beyond the number of concentration states $n_c-3$ when this synthesis process is used.

Recently, circuits have been developed for specific charging conditions such as charge relaxation \cite{torregrossa2014improvement}. With the proposed PB approach, there should be a much broader range of dynamics which are considered in forming the impedance functions. This should result in more generalised synthesized circuits that are designed for a broader range of charging conditions. The ability to realise the physical dynamics in terms of both the \textit{ladder} and the \textit{dynamic} circuits contrasts with the view in the literature which gives each of these circuits a distinct physical interpretation; the \textit{ladder} circuit is said to describe ion movement down a pore while the \textit{dynamic} circuit is said to model double layer effects \cite{zhang2015comparative}. The analysis of this paper implies that the physical interpretation of these circuits is not as well defined as that. The inversion of the impedance function $G(s)$ is known as the admittance function $F(s)$ which is the frequency space voltage to current gain and can also be synthesized in terms of resistors and inductors \cite{wadhwa2007network}. In the proposed method, the PB model is transformed into a circuit and it is believed that this transformation can not occur in the opposite direction, i.e. going from the circuit equations to the model PDEs. However, it is pointed out that developing PB models by considering the physics would be a better approach than by reverse synthesizing circuits.

The proposed circuit synthesis method is flexible since it can be applied to general PB models describing additional physical phenomena. For example, including the linear dependence of electrolyte conductivity with concentration
\begin{align}
\kappa = \kappa_0 c,
\end{align}
which is mentioned, but not enforced, in \cite{verbrugge2005microstructural}, changes the state-space system to
\begin{align}\label{system_conductivity}
\begin{split}
&\begin{bmatrix}\epsilon & \frac{aC}{F}(t_-\frac{dq_+}{dq}+t_+\frac{dq_-}{dq}) & 0\\
0 & aC & 0 \\
0 & 0 & 0 \end{bmatrix}
\begin{bmatrix} \dot{c} \\ \dot{\phi}_1-\dot{\phi}_2 \\ \dot{\phi}_2\end{bmatrix}
= 
 \begin{bmatrix}D \frac{\partial^2}{\partial x^2} & 0 & 0\\
0 &\sigma \frac{\partial^2}{\partial x^2}&\sigma \frac{\partial^2}{\partial x^2} \\
\frac{\kappa_0 RT(t_+-t_-)}{F} \frac{\partial}{\partial x}& \sigma \frac{\partial}{\partial x} & \kappa_0 c \frac{\partial}{\partial x}+ \sigma \frac{\partial}{\partial x} \end{bmatrix}
\begin{bmatrix} c \\ \phi_1- \phi_2 \\ \phi_2\end{bmatrix} 
+\begin{bmatrix}0 \\0 \\ 1 \end{bmatrix}i.
\end{split}
\end{align}
whose linearised impedance is
\begin{align}\label{impedance_kappa}
G(s) = \frac{V(s)}{I(s)} = \frac{(s+4.76)(s+0.3)(s+2.95 \times 10^{-5})(s+3.27 \times 10^{-5})} {s(s+3.74)(s+2.9 \times 10^{-5})(s+3.27 \times 10^{-6}) }.
\end{align}
As well as electrolyte conductivity, the capacitance $aC$ has also been shown to vary during charging. Several models have been proposed to account for this relationship, with the most popular being the Guoy-Chapman-Stern model \cite{lu2013supercapacitors}. In the region of low voltage, this relationship can be approximated by a linear fit
\begin{align}
aC = \alpha + \beta (\phi_1-\phi_2).
\end{align}
 This relationship changes the state-space system to
\begin{align}\label{system_capacitance}
\begin{split}
&\begin{bmatrix}\epsilon & \frac{\alpha + \beta (\phi_1-\phi_2)}{F}(t_-\frac{dq_+}{dq}+t_+\frac{dq_-}{dq}) & 0\\
0 & \alpha + \beta (\phi_1-\phi_2) & 0 \\
0 & 0 & 0 \end{bmatrix}
\begin{bmatrix} \dot{c} \\ \dot{\phi}_1-\dot{\phi}_2 \\ \dot{\phi}_2\end{bmatrix}
= 
\begin{bmatrix} 0 \\0 \\ \kappa \bigg(\frac{RT(t_+-t_-)}{F}\bigg) \frac{\partial}{\partial x}\end{bmatrix}\text{ln }(c) 
\\
& \qquad + \begin{bmatrix}D \frac{\partial^2}{\partial x^2} & 0 & 0\\
0 &\sigma \frac{\partial^2}{\partial x^2}&\sigma \frac{\partial^2}{\partial x^2} \\
0& \sigma \frac{\partial}{\partial x} & \kappa \frac{\partial}{\partial x}+ \sigma \frac{\partial}{\partial x} \end{bmatrix}
\begin{bmatrix} c \\ \phi_1- \phi_2 \\ \phi_2\end{bmatrix} 
+\begin{bmatrix}0 \\0 \\ 1 \end{bmatrix}i.
\end{split}
\end{align}
and the impedance for a given $\alpha$ and $\beta$ in Table \ref{tab:GlobalParams} to
\begin{align}\label{impedance_ac}
G(s) = \frac{V(s)}{I(s)} = \frac{s^2(s+6.04)(s+1.46)(s+0.27)(s+0.0031)}  {s^3(s+5.17)(s+1.29)(s+0.003) }.
\end{align}
Both impedance functions (\ref{impedance_kappa}) \& (\ref{impedance_ac}) are positive real functions and so can be synthesized by passive circuit elements. 
The method can also be easily updated to generate local circuits applicable to any operating region besides the equilibrium concentration by changing the linearisation point of the model. 

\begin{figure}
\centering
\graphicspath{ {./Figures/} }
\includegraphics[width=0.75\columnwidth]{{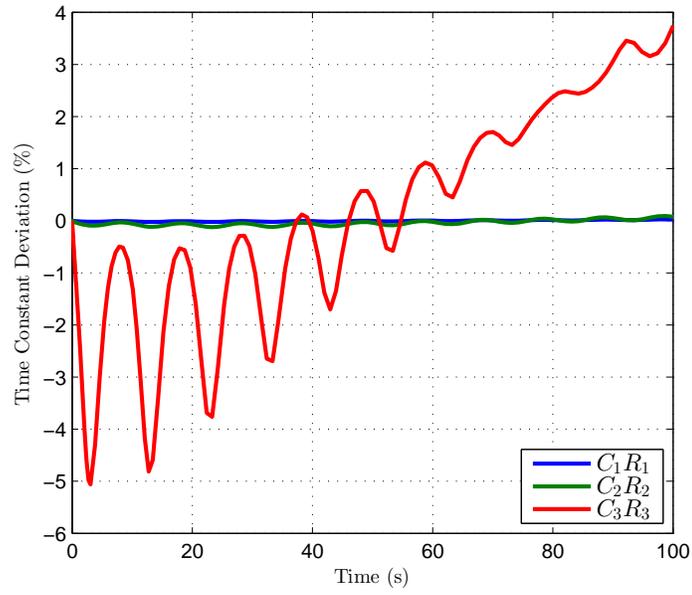}}
\caption{Variation of the time constants of the dynamic circuit during a charge with sinusoidal current.}
\label{fig:params}
\end{figure}

As discussed in the introduction, circuits describe the local dynamics of a supercapacitor in some operating region. During a charging profile, the supercapacitor may leave this region, necessiating the need for a new circuit to be generated, typically using parameter estimaton methods \cite{zhang2014online}. This parameter esimtation problem for the circuit components can be re-cast as a synthesis problem of the PB model. Fig \ref{fig:params} shows the percentage deviation from their original values of the three time constants of the dynamic circuit (\ref{dynamic}) that were synthesized from (\ref{state_space}) with an input current of $i =10 + 10 \sin(0.1 t)$ A m$^{-2}$ where $t$ is time.  It was found that only the time constant $R_3C_3$ related to long term dynamics changed significantly during this charging profile. 
This is because the localised component of the PB model  dynamics that changes during the charging profile is due to the nonlinearity which is solely a function of the concentration. Since the concentration exhibits slow diffusion dynamics, only the long term time constants are local. 
It is noted that if online parameter estimation of the circuit was to be implemented using this approach, it would require the (higher-order) PB model to be ran simultaneously. This would limit the applicability of the approach in practise, however, it gives a physical interpretation of the circuit parameters during a charge which would not be possible using parameter esitmation methods.

\section{Conclusion}
In this paper, a method was proposed for synthesising electrical circuits from physical supercapacitor models. This method used model discretisation, linearisation, balanced model order reduction and passive network synthesis. The method is flexible since a wide class of circuits can be realised from a wide class of physical models. The circuits were validated by comparing their Bode responses to that of a linearised physical model. The aim of this paper is to give a greater understanding to the physical interpretation of equivalent circuit models and also to enable the synthesis of more general circuits whose impedance function would be generated by considering the device physics, not by experimental fitting. 


\newpage
\bibliographystyle{unsrt} 
\bibliography{bibliog}
\newpage

\begin{table} [h!]
\centering 
\renewcommand{\arraystretch}{1.3}
\begin{tabular}{|c| c| }
\hline 
\multicolumn{2}{|c|}{Classical Model} \\
\hline
Component & Value\\
\hline
$R_1$ & $6.2  \times 10^{-4} ~ \Omega$ \\
\hline
$R$ & $ 2.5 \times 10^{-3} ~ \Omega$ \\
\hline
$C_1$ & $ 431 ~  F $ \\
\hline
$C$ & $1.05 \times 10^{3} ~ F$  \\
\hline 
\end{tabular}
\caption{Resistors and capacitors of the classical circuit.}
\label{tab:Classical Model}
\end{table}

\begin{table} [h!]
\centering 
\renewcommand{\arraystretch}{1.3}
\begin{tabular}{|c| c| }
\hline 
\multicolumn{2}{|c|}{Dynamic Model} \\
\hline
Component & Value\\
\hline
$R_1$ & $ 3.75 \times 10^{-4} ~ \Omega$ \\
\hline
$R_2$ &$ 3.15 \times 10^{-4} ~\Omega$ \\
\hline
$R_3$ &$  2.52 \times 10^{-4} ~\Omega$ \\
\hline
$R$ &$ 2.52 \times 10^{-3} ~\Omega$ \\
\hline
$C_1$ &475 $F$ \\
\hline
$C_2$ & $2.26 \times 10^{3} ~ F $ \\
\hline
$C_3$ & $ 1.3 \times 10^{6} ~ F$ \\
\hline
$C$ & $ 1.05 \times 10^{3} ~ F$ \\
\hline 
\end{tabular}
\caption{Resistors and capacitors of the dynamic circuit.}
\label{tab:Dynamic Model}
\end{table}

\begin{table} [h!]
\centering 
\renewcommand{\arraystretch}{1.3}
\begin{tabular}{|c| c| }
\hline 
\multicolumn{2}{|c|}{Ladder Model} \\
\hline
Component & Value\\
\hline
$R_1$ &  $2.5 \times 10^{-3} ~ \Omega$ \\
\hline
$R_2$ & $9.8 \times 10^{-4} ~ \Omega$ \\
\hline
$R_{3}$ &  $4 \times 10^{-3} ~ \Omega$  \\
\hline
$C_1$ & $285.9 ~ F$\\
\hline
$C_2$ & $549.6 ~ F $ \\
\hline
$C_3$ & $249.6 \times  10^{-6} ~ F$ \\
\hline 
\end{tabular}
\caption{Resistors and capacitors of the ladder circuit.}
\label{tab:Ladder Model}
\end{table}

\begin{table} [h!]
\centering 
\renewcommand{\arraystretch}{1.3}
\begin{tabular}{|c| c| c|}
\hline
Parameter &Value & Units \\
\hline 
\multicolumn{3}{|c|}{Global Parameters} \\
\hline
$\frac{dq_+}{dq}= \frac{dq_-}{dq}$ & -0.5 & \\
$t_+$ & 0.55&  \\
$T$ & 298 & K \\
\hline
\multicolumn{3}{|c|}{Electrode Parameters} \\
\hline
$\kappa$ &0.0195& S m$^{-1}$ \\
$D$ & 2.09 $\times 10^{-12}$ & m$^2$ s$^{-1}$  \\
$\epsilon$ & 0.67 & \\
$\sigma$ & 0.0521 & S m$^{-1}$ \\
$aC$ & 42 $\times 10^6$ & F m$^{-2}$ \\
$\alpha$ & 42 $\times 10^6$ & F m$^{-2}$ \\
$\beta$ & 10 $\times 10^6$ & F m$^{-2}$ \\
\hline 
\multicolumn{3}{|c|}{Separator Parameters} \\
\hline
$\kappa$ &0.0312& S m$^{-1}$ \\
$D$ & 3.34 $\times 10^{-12}$ & m$^2$ s$^{-1}$ \\
$\epsilon$ & 0.6 & \\
\hline 
\end{tabular}
\caption{Parameters for the PB supercapacitor model \cite{verbrugge2005microstructural}.}
\label{tab:GlobalParams}
\end{table}

\end{document}